\documentstyle[11pt,moriond,epsfig]{article}

\def\pr#1{#1^\prime}

\def\beq{\begin{equation}}
\def\eeq{\end{equation}}

\def\beqn{\begin{eqnarray}}
\def\eeqn{\end{eqnarray}}
\relax

\def\s{\sigma}
\def\fix{\right|_{\rm FO}\!\!\!\!}
\def\res{\right|_{\rm res}\!\!\!\!}
\def\imp{\right|_{\rm imp}\!\!\!\!}
\def\TNLL{\right|_{\rm TNLL}\!\!}
\def\NLL{\right|_{\rm NLL}\!\!\!\!}
\def\LL{\right|_{\rm LL}\!\!}

\def\lq{\left[}
\def\rq{\right]}

\def\({\left(}
\def\){\right)}
\def\xe{x_E}
\def\xp{x_p}
\def\zp{z_p}

\newcommand\LambdaQCD{\Lambda_{\scriptscriptstyle \rm QCD}}

\def    \be             {\begin{equation}}
\def    \ee             {\end{equation}}
\def    \ba             {\begin{eqnarray}}
\def    \ea             {\end{eqnarray}}

\def    \=              {\;=\;}
\def    \frac           #1#2{{#1 \over #2}}

\def \ep{\epsilon}
\def \as   {\ifmmode \alpha_s \else $\alpha_s$ \fi}

\def\b0{b_0}

\def \mt   {\ifmmode m_{\rm t} \else $m_{\rm t}$ \fi}

\def \to   {\mbox{$\rightarrow$}}
\newcommand     \MSB            {\ifmmode {\overline{\rm MS}} \else
                                 $\overline{\rm MS}$\fi}

\newcommand\hepph[1]{{\tt hep-ph/#1}}

\newcommand     \epem           {\ifmmode{e^+e^-}\else{$e^+e^-$}\fi}
\def \to   {\mbox{$\rightarrow$}}

\def\figura#1#2#3
{\begin{figure}[hb]
  \begin{center}
  \leavevmode\protect\epsfxsize=#1\protect\epsffile{#2.eps}
  \protect\caption{#3}
  \label{#2}
  \end{center}
  \end{figure}
}
\newlength{\Largfig}
\def\({\left(}
\def\){\right)}

\def\s{\sigma}

\def\asb{{}\ifmmode \bar{\alpha}_s \else $\bar{\alpha}_s$\fi}

\def\Dnp{D_{\rm NP}}

\def\OPAL{\mbox{OPAL}}
\def\ALEPH{\mbox{ALEPH}}
\def\ARGUS{\mbox{ARGUS}}

\def\ord#1{{\cal O}\(#1\)}

\relax
\def\pl#1#2#3{{\it Phys. Lett. }{\bf #1}\ (19#2)\ #3}

\def\prl#1#2#3{{\it Phys. Rev. Lett. }{\bf #1}\ (19#2)\ #3}
\def\rmp#1#2#3{{\it Rev. Mod. Phys. }{\bf#1}\ (19#2)\ #3}
\def\prep#1#2#3{{\it Phys. Rep. }{\bf #1}\ (19#2)\ #3}
\def\pr#1#2#3{{\it Phys. Rev. }{\bf #1}\ (19#2)\ #3}
\def\np#1#2#3{{\it Nucl. Phys. }{\bf #1}\ (19#2)\ #3}
\def\sjnp#1#2#3{{\it Sov. J. Nucl. Phys. }{\bf #1}\ (19#2)\ #3}
\def\app#1#2#3{{\it Acta Phys. Polon. }{\bf #1}\ (19#2)\ #3}
\def\jmp#1#2#3{{\it J. Math. Phys. }{\bf #1}\ (19#2)\ #3}
\def\nc#1#2#3{{\it Nuovo Cim. }{\bf #1}\ (19#2)\ #3}

\relax

\begin{document}

\nopagebreak
{\flushright{
        \begin{minipage}{4cm}
        DTP/99/46 \\
        {\tt hep-ph/9905344}\hfill \\
         May 1999\\ 
        \end{minipage}        }

}
\vspace*{3cm}
\title{ON THE EXTRACTION OF NON-PERTURBATIVE EFFECTS IN THE FRAGMENTATION
  FUNCTIONS OF HEAVY QUARKS IN ${ e^+e^-}$ ANNIHILATION
\footnote{Talk presented at {\sl XXXIV Rencontres De Moriond}, Les Arcs
  1800, France, 20--27 March 1999. To appear in the proceedings of: ``QCD and
  High Energy Hadronic Interactions''.}}

\author{CARLO OLEARI}

\address{Department of Physics, University of Durham, Durham, DH1 3LE, UK}

\maketitle
\abstracts{  
  In this talk, we consider the computation of $D$ and $B$ fragmentation
  functions in $\epem$ annihilation. We present an improved differential
  cross section that merges together the ${\cal O}(\as^2)$ fixed-order
  calculation and the next-to-leading-logarithmic resummed one.  We
  compare the  
  results of fitting present data using the next-to-leading-log
  cross section, the ${\cal O}(\as^2)$ fixed-order and the
  improved one. }

\section{Introduction}
The interest in refining our knowledge of the fragmentation
functions for heavy quarks (HQFF) in $\epem$ annihilation stems mainly from the
possibility 
of using them to improve
our understanding of heavy-flavour hadroproduction and photoproduction.
In fact, this is a quite clean environment where the impact of
non-perturbative effects can be studied.

The aim of this talk is to present new results on the size of
non-perturbative effects on heavy-flavour production in $\epem$ annihilation.

The outline of the talk is as follows. We first give a brief introduction of
the fragmentation-function formalism and we introduce an improved
differential cross section, which merges together the $\ord{\as^2}$ fixed-order
result with the next-to-leading-logarithmic resummed one.
We parametrize the non-perturbative contributions to the differential cross
section using the Peterson form~\cite{Peterson}.
We present some fits to recent data from which we extract the
non-perturbative part. Then we conclude with some comments on the
obtained results.

\section{Theoretical framework}
\label{sec:Theoretical}
We consider the inclusive production of a heavy quark
$Q$ of mass $m$, in the process
\beq 
\label{eq:process}
  e^+ e^- \,\to\, Z/\gamma\;(q) \,\to\, Q\,(p) + X\;,
\eeq
where $q$ and
$p$ are the four-momenta of the intermediate boson and of the final quark.
We define $\xe$ the scaled energy of the final heavy quark, and $\xp$
the normalized momentum fraction 
\beq
   \xe = \frac{2\, p\cdot q}{q^2}\;, \quad \quad 
 \xp = \frac{\sqrt{\xe^2 - \rho}}{\sqrt{1-\rho}}\;,\quad\quad
  {\rm where:} \quad \rho= \frac{4\,m^2}{E^2} 
   \quad{\rm and} \quad  E=\sqrt{q^2}\;.
\eeq
The inclusive cross section for the production of a heavy quark 
can be written as a perturbative expansion in $\as$
\beq
\label{eq:sigma}
   \frac{d\sigma}{d\xp}
= \sum_{n=0}^{\infty} a^{(n)}(\xp,E,m,\mu)\, \asb^n(\mu)\;,
\quad\quad\quad \asb(\mu) =  \frac{\as(\mu)}{2\pi}\;,
\eeq
where $\mu$ is the renormalization scale.

If $\mu \approx E \approx m$, the truncation of Eq.~(\ref{eq:sigma}) 
at some fixed
order in the coupling constant can be used to approximate the cross
section.  An $\ord{\as^2}$ fixed-order calculation for the
process~(\ref{eq:process}) is
available~\cite{zbb4,Rodrigo,Bernreuther}, so that we can compute the
coefficients of Eq.~(\ref{eq:sigma}) at the $\ord{\as^2}$ level.
We  define the fixed-order (FO) result as
\beq
\label{eq:sigma_fix}
 \left.  \frac{d\sigma}{d\xp}\fix
= a^{(0)}(\xp,E,m) + a^{(1)}(\xp,E,m)\, \asb + a^{(2)}(\xp,E,m)\,
 \asb^2 \;,
\eeq
where we have taken $\mu=E$ for ease of notation, and, from now on,
$\asb\ \equiv \left.\asb(\mu)\right|_{\mu=E}$. 

If $E\gg m$, large logarithms of the form $\log\(E^2/m^2\) $ appear
in the differential cross section~(\ref{eq:sigma}) to
all orders in $\as$. In this limit, if we disregard all
power-suppressed terms of the form $m^2/E^2$,
the inclusive cross section can be organized in the expansion
\beq
\label{eq:resummed_sigma}
  \left. \frac{d\sigma}{dx}\res =
\sum_{n=0}^{\infty} \beta^{(n)}(x) \,\( \asb L
\)^n \!+ \asb\sum_{n=0}^{\infty} \gamma^{(n)}(x) \,\( \asb
L\)^n
 + \asb^2  \sum_{n=0}^{\infty}  \delta^{(n)}(x) \,
 \( \asb L\)^n + 
\ldots \;,
\eeq
where $L=\log\(E^2/m^2\)$ and $x$ stands for either $\xe$ or $\xp$,
since the two variables 
differ by power-suppressed effects.

The leading-logarithmic (LL) and next-to-leading-logarithmic (NLL) 
cross sections are given, respectively, by
\beq
\label{eq:NLL}
\left. \frac{d\sigma}{dx} \LL  =
\sum_{n=0}^{\infty} \beta^{(n)}(x) \,\( \asb L
\)^n\;,  \quad\quad
  \left. \frac{d\sigma}{dx} \NLL =
\sum_{n=0}^{\infty} \beta^{(n)}(x) \,\( \asb L
\)^n \!+ \asb\sum_{n=0}^{\infty} \gamma^{(n)}(x) \,\( \asb
L\)^n\,.
\eeq
The $\delta^{(n)}$ coefficients of Eq.~(\ref{eq:resummed_sigma}) define the
NNLL terms, that are, as of 
now, not known. The technique to resum these large logarithms is well
established and more details can be found in Ref.~\cite{MeleNason}.

The expansion of the NLL differential cross section up to order $\as^2$, that
we call 
the truncated NLL (TNLL from now on), is given by
\beq
\left. \frac{d\sigma}{dx} \TNLL 
=\beta^{(0)}(x) + \asb \( \gamma^{(0)}(x) + \beta^{(1)}(x)\,
L\) + \asb^2 \( \gamma^{(1)}(x)\,L + 
\beta^{(2)}(x)\, L^2 \) \;,
\eeq
and does not coincide with the massless limit of the FO calculation:
a term of order $\as^2$, not accompanied by logarithmic factors,
may in fact survive in the massless limit of the FO result.  In
the HQFF approach, this is a NNLL effect, and therefore it is not included
at the NLL level. 

It is now clear how to obtain an improved formula, which contains all the
information present in the FO approach, as well as in the HQFF approach.
Using Eqs.~(\ref{eq:sigma_fix}) and~(\ref{eq:NLL}), we write the
improved cross section as 
\beq
\label{eq:improved}
   \left. \frac{d\sigma}{d\xp}\imp = 
\sum_{i=0}^2
a^{(i)}(\xp,E,m)\,\asb^i
 +\sum_{n=3}^{\infty} \beta^{(n)}(x) \,\( \asb L
\)^n  +\asb\sum_{n=2}^{\infty} \gamma^{(n)}(x) \,\( \asb
L\)^n \;,
\eeq
where the LL and NLL sums now start from $n=3$ and $n=2$
respectively, in order to avoid double counting.
Formula (\ref{eq:improved}) includes exactly all
terms up to the order $\as^2$ with mass effects, and all terms of the
form $\Big( \asb L \Big)^n $ and $\asb \, \Big( \asb
L \Big)^n $, so that it is also
correct at NLL level for $E \gg m$. It can be viewed as an
interpolating formula: for moderate energies, it is accurate to the order
$\as^2$, while for very large energies it is accurate at the NLL level.

\section{Non-perturbative effects}
\label{sec:NonPerturbative}

When dealing with the HQFF formalism, we have to face the
problem that the initial condition for the evolution of the fragmentation
functions, which is computed as a power expansion in terms of $\as(m)$,
contains irreducible, non-perturbative uncertainties of order
$\LambdaQCD/m$~\cite{MeleNason}. 
In addition, we need a model to describe the hadronization phenomena, that is
the heavy quark turning into a heavy-flavoured hadron.  We assume that all
these effects, together with all low-energy ones, are described by a
non-perturbative 
fragmentation function $\Dnp^H$, so that we can write the full hadronic cross
section $d\s^H/dx$, 
including non-perturbative corrections, as
\beq
\label{eq:hadfactor}
   \frac{d\s^H}{dx} = \frac{d\s^P}{dx}\otimes \Dnp^H (x) \;,
\eeq
where $d\s^P/dx$ is the partonic differential cross section.

We will parametrize the non-perturbative part of the
fragmentation function with the one-parameter-dependent Peterson
form~\cite{Peterson} 
\beq
\label{eq:peterson}
\Dnp^H (x) = P(x,\ep) \equiv 
 N \,\frac{x\,(1-x)^2}{\lq (1-x)^2+x\,\ep\rq^2} \;,
\quad\quad {\rm with: } \quad 
\sum_H  \int_0^1  dx \, \Dnp^H (x) = 1\;,
\eeq
where the normalization factor $N$ determines
the fraction of the hadron of type $H$ in the final state.

\section{Fit to experimental data}
\label{sec:Fits}

By fitting the experimental data with the differential cross section of
Eq.~(\ref{eq:hadfactor}), we can extract the
non-perturbative part of the fragmentation functions.

We consider the data sets for $D$ production obtained by \ARGUS~\cite{ARGUS},
at $E=10.6$~GeV, and \OPAL~\cite{OPAL}, at $E=91.2$~GeV, and the data for $B$
production obtained by \ALEPH~\cite{ALEPH}, at $E=91.2$~GeV.

Besides the LL and NLL fits (similar to that ones made in
Ref.~\cite{CacGre}), we present new fits with our improved cross
section.

We have fitted the data by $\chi^2$ minimization, allowing 
both the value of $\ep$ and the normalization to float.
We have kept  $\LambdaQCD^{(5)}$ fixed to $200$~MeV.
The results of the fits are displayed in Table~\ref{tab:ep_values}.
\begin{table}[htbp]
 \leavevmode
\small
%\footnotesize
    \caption{Results of the fit of the non-perturbative $\ep$
 parameter for the Peterson fragmentation function. The value of
 $\chi^2$/dof is given in parenthesis. The range of the fit is
 indicated in Figure~1 with small crosses.
$\mbox{}^{(*)}\!\!$ We have excluded the first three experimental
  points for 
 the $\ord{\as}$ and $\ord{\as^2}$ fixed-order fit to the \OPAL\
 data. \hfill\mbox{}} 
\vspace{0.2cm}
 \begin{center}
\begin{tabular}{c||c|c|c|c|c}
 \hline\hline
$\ep\;(\chi^2/{\rm dof})$
& $\as$ fixed order & LL &
$\as^2$ fixed order  & NLL  & NLL improved \\
 \hline\hline
ARGUS $D$ & 0.058 (0.852) & 0.053 (2.033) &  0.035 (0.855)&
0.018 (1.234) & 0.022 (1.210) \\
OPAL $D$ &  0.078 $\!\!\mbox{}^{(*)}\!\!$ (0.706) & 0.048 (1.008)  
& 0.040 $\!\!\mbox{}^{(*)}\!\!$ (0.769) & 0.016 (1.122) 
& 0.019 (1.066)\\
ALEPH $B$ & 0.0069 (4.607)  & 0.0061 (0.137)  &  0.0033
(2.756) & 0.0016 (0.441)  & 0.0023 (0.635) \\
 \hline\hline
    \end{tabular}
  \label{tab:ep_values}
  \end{center}
\end{table}
The corresponding curves, together with the data, are shown
in Figure~1.
%%%%%%%%%%%%%%%%%%%%%%%%%%%%%%%%%%%%%%%%%%%
\begin{figure}[htb]
\label{fig:fit}
\centerline{\epsfig{figure=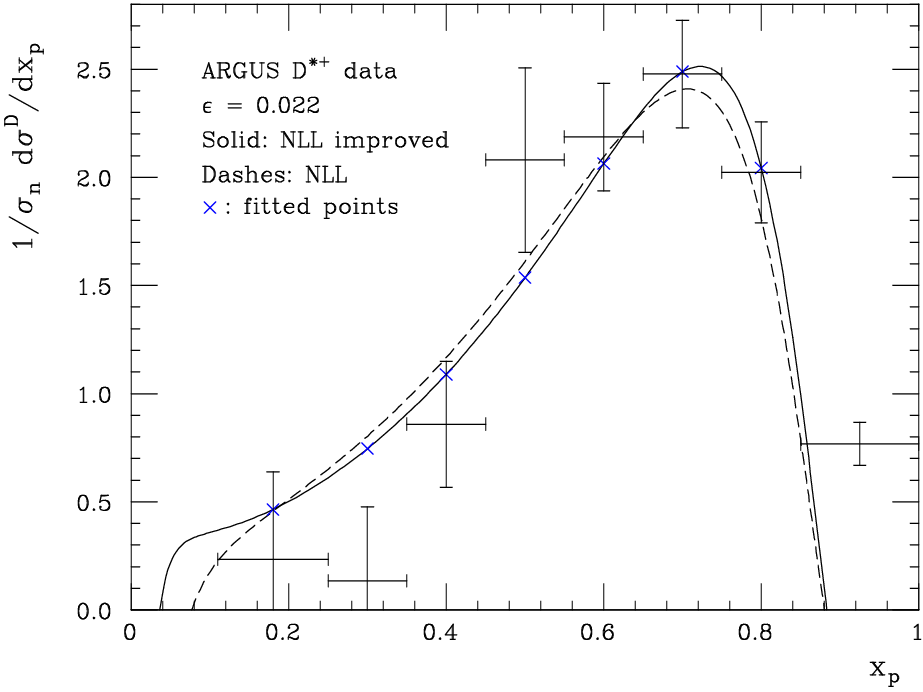,width=0.45\textwidth,clip=}
\epsfig{figure=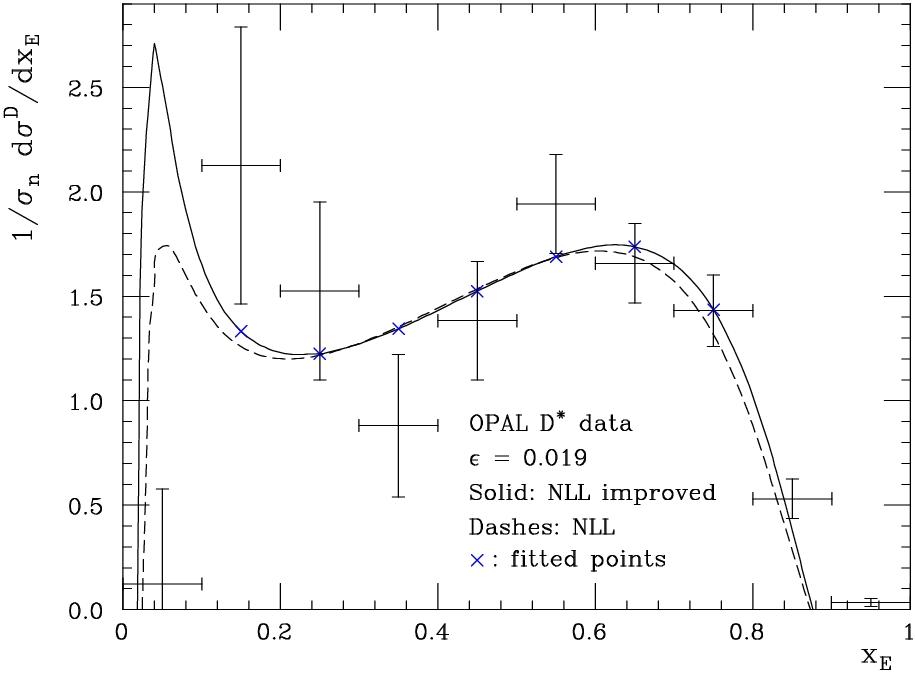,width=0.45\textwidth,clip=} 
}
\vspace{3mm}
\centerline{
\epsfig{figure=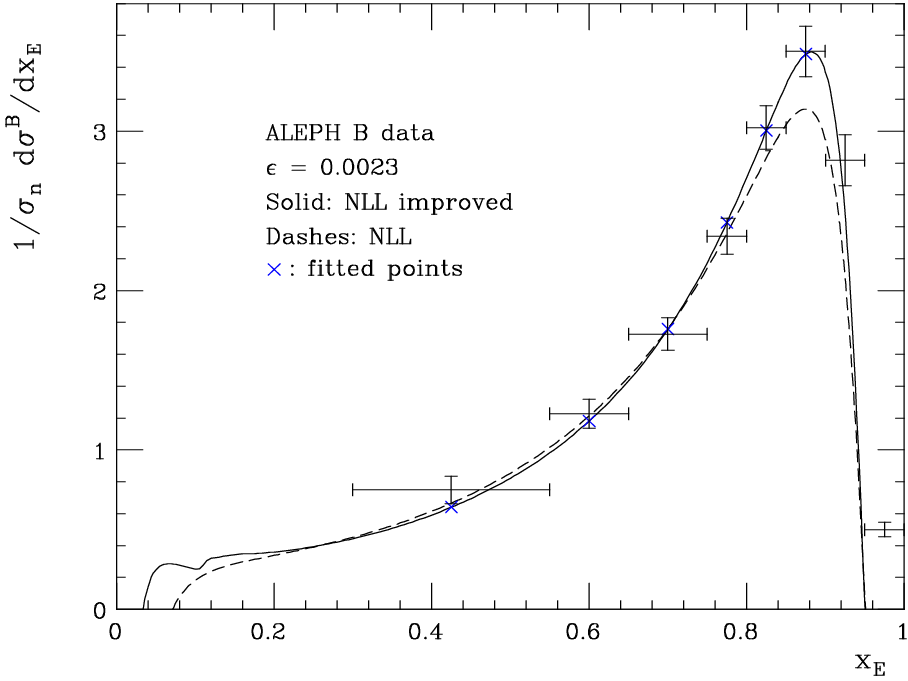,width=0.45\textwidth,clip=}
\begin{minipage}[b]{2 mm}
\mbox{}
\end{minipage}
\begin{minipage}[b]{0.43\textwidth}
\footnotesize
Figure 1: Best fit for the improved fragmentation function 
at \ARGUS\ (up left), \OPAL\ (up right) and \ALEPH\ (left).  
In dashed line, the NLL fragmentation function at the same value of $\ep$.
The overall normalization factor $\sigma_{\rm n}$ of the curves is immaterial,
because of the uncertainty in the specific heavy-flavoured hadron fraction,
and it must be fitted to the data. It should be, however, finite
in the massless limit~\cite{frag99}.
\vspace{1.3cm}
\end{minipage}
}
\end{figure}
%%%%%%%%%%%%%%%%%%%%%%%%%%%%%%%%%%%%%%%%%%%
The full improved resummed result of Eq.~(\ref{eq:improved}) has been used
here.  For comparison, we have also plotted the NLL curves computed at the
same value of $\ep$.
The small differences we find in the LL and NLL sectors, with respect to
the results of Ref.~\cite{CacGre},
are due to different range, normalization and
adjustment of physical parameters.

\section{Conclusions}
\label{sec:conclusions}
In this talk we have presented the results obtained using an improved
differential cross section, that contains leading and next-to-leading
logarithmic resummation, fixed-order effects up to $\ord{\as^2}$, and mass
effects to the same order, in the fits of present data for
heavy-flavour fragmentation functions in $\epem$ collisions.
Our finding can be easily summarized as follows.  We generally find little
differences between our results and the NLL resummed calculation. This
indicates that mass effects are of limited importance in
fragmentation-function physics in $\epem$ annihilation.  On the other
hand, our calculation confirms the fact that, when NLL effects are
included, the importance of a non-perturbative initial condition is
reduced. 

\section*{Acknowledgments}
The results shown here have been obtained in collaboration with Paolo
Nason~\cite{frag99,frag98-lett}. 

\relax
\def\pl#1#2#3{{\it Phys.\ Lett.\ }{\bf #1}\ (19#2)\ #3}
\def\zp#1#2#3{{\it Z.\ Phys.\ }{\bf #1}\ (19#2)\ #3}
\def\prl#1#2#3{{\it Phys.\ Rev.\ Lett.\ }{\bf #1}\ (19#2)\ #3}
\def\rmp#1#2#3{{\it Rev.\ Mod.\ Phys.\ }{\bf#1}\ (19#2)\ #3}
\def\prep#1#2#3{{\it Phys.\ Rep.\ }{\bf #1}\ (19#2)\ #3}
\def\pr#1#2#3{{\it Phys.\ Rev.\ }{\bf #1}\ (19#2)\ #3}
\def\np#1#2#3{{\it Nucl.\ Phys.\ }{\bf #1}\ (19#2)\ #3}
\def\sjnp#1#2#3{{\it Sov.\ J.\ Nucl.\ Phys.\ }{\bf #1}\ (19#2)\ #3}
\def\app#1#2#3{{\it Acta Phys.\ Polon.\ }{\bf #1}\ (19#2)\ #3}
\def\jmp#1#2#3{{\it J.\ Math.\ Phys.\ }{\bf #1}\ (19#2)\ #3}
\def\nc#1#2#3{{\it Nuovo Cim.\ }{\bf #1}\ (19#2)\ #3}
\def\jhep#1#2#3{{\it J.\ High Energy Phys.\ }{\bf #1}\ (19#2)\ #3}
\relax


\section*{References}
\begin{thebibliography}{99}
\bibitem{Peterson}
C.~Peterson, D.~Schlatter, I.~Schmitt and P.~M.~Zerwas, \pr{D27}{83}{105}.
\bibitem{zbb4}
P.~Nason and C.~Oleari, \np{B521}{98}{237};\newline
C.~Oleari, Ph.~D.~Thesis, \hepph{9802431}.
\bibitem{Rodrigo}
  G. Rodrigo, {\it Nucl. Phys. Proc. Suppl.} {\bf 54A}\ (1997)\ 60; \newline
  G. Rodrigo, Ph. D. Thesis,
  Univ. of Val\`encia, 1996, \hepph{9703359};\newline
  G.~Rodrigo, A.~Santamaria and M.~Bilenkii, \prl{79}{97}{193}.
\bibitem{Bernreuther}
  W. Bernreuther, A. Brandenburg and P. Uwer, \prl{79}{97}{189};
 A. Brandenburg and P. Uwer, \np{B515}{98}{279}.
\bibitem{MeleNason}
B.~Mele and P.~Nason, \np{B361}{91}{626}.
\bibitem{ARGUS}
H.~Albrecht et al., ARGUS Collaboration, \zp{C52}{91}{353}.
\bibitem{OPAL}
R.~Akers et al., OPAL Collaboration, \zp{C67}{95}{27}.
\bibitem{ALEPH}
D.~Buskulic et al., ALEPH Collaboration, \pl{B357}{95}{699}.
\bibitem{CacGre}
M.~Cacciari and M.~Greco, \pr{D55}{97}{7134}.
\bibitem{frag99}
P.~Nason and C.~Oleari, \hepph{9903541}, submitted to {\it Nucl.~Phys.\ }{\bf
  B}. 
\bibitem{frag98-lett}
P.~Nason and C.~Oleari, \pl{B447}{99}{327}.
\end{thebibliography}
\end{document}